\documentclass[amsfonts, amssymb, amsmath, reprint, showkeys, nofootinbib]{revtex4-1}
\usepackage[english]{babel}
\usepackage[utf8]{inputenc}
\usepackage[colorinlistoftodos, color=green!40, prependcaption]{todonotes}
\usepackage{amsthm}
\usepackage{mathtools}
\usepackage{physics}
\usepackage{xcolor}
\usepackage{graphicx}
\usepackage[left=23mm,right=13mm,top=35mm,columnsep=15pt]{geometry} 
\usepackage{adjustbox}
\usepackage{placeins}
\usepackage[T1]{fontenc}
\usepackage{lipsum}
\usepackage{csquotes}
\usepackage[pdftex, pdftitle={Article}, pdfauthor={Author}]{hyperref} 

\newcommand{\noopsort}[1]{}

\begin{document}
\title{Deep-learning Optical Flow Outperforms PIV in\\Obtaining Velocity Fields from Active Nematics}

\author{Phu N. Tran,$^{a}$ Sattvic Ray,$^{c}$ Linnea Lemma,$^{a,c}$ Yunrui Li,$^{b}$ Reef Sweeney,$^{c}$ Aparna Baskaran,$^{a}$ Zvonimir Dogic,$^{a,c,d}$ Pengyu Hong,$^{b\ast}$ Michael F. Hagan$^{a\ddag}$ \vspace{.5cm}}

    \affiliation{$^{a}$Department of Physics, Brandeis University, Waltham, MA 02453, USA}
    \affiliation{$^{b}$Department of Computer Science, Brandeis University, Waltham, MA 02453, USA}
    \affiliation{$^{c}$Department of Physics, University of California at Santa Barbara, Santa Barbara, CA 93106, USA}
    \affiliation{$^{d}$Biomolecular and Engineering Science, University of California at Santa Barbara, Santa Barbara, CA 93106, USA}

\author{Correspondence email addresses:\\$^{\ast}$hongpeng@brandeis.edu, $^{\ddag}$hagan@brandeis.edu \vspace{.5cm}}

\date{\today} 

\begin{abstract}
Deep learning-based optical flow (DLOF) extracts features in adjacent video frames with deep convolutional neural networks. It uses those features to estimate the inter-frame motions of objects at the pixel level. In this article, we evaluate the ability of optical flow to quantify the spontaneous flows of MT-based active nematics under different labeling conditions. We compare DLOF against the commonly used technique, particle imaging velocimetry (PIV). We obtain flow velocity ground truths either by performing semi-automated particle tracking on samples with sparsely labeled filaments, or from passive tracer beads. We find that DLOF produces significantly more accurate velocity fields than PIV for densely labeled samples. We show that the breakdown of PIV arises because the algorithm cannot reliably distinguish contrast variations at high densities, particularly in directions parallel to the nematic director. DLOF overcomes this limitation. For sparsely labeled samples, DLOF and PIV produce results with similar accuracy, but DLOF gives higher-resolution fields. Our work establishes DLOF as a versatile tool for measuring fluid flows in a broad class of active, soft, and biophysical systems. 
(Data and code is available at https://github.com/tranngocphu/opticalflow-activenematics)
\end{abstract}


\maketitle

\section{Introduction}

Accurate measurement of flow fields is a cornerstone for modeling diverse phenomena that range across the field of fluid dynamics \citep{corpetti2002}, active matter \citep{marchetti2013}, and biological systems \citep{vogel2020}. A conventional approach to estimating flow fields is Particle Image Velocimetry (PIV), where flow velocities are computed by correlating features of two consequent images~\citep{thielicke2021,sarno2018,raffel2018}. However, PIV has limitations. One arises from the dependence of the interrogation window size on seeding particle speed. Consequently, PIV cannot estimate turbulent flows smaller than the interrogation window, leading to potential errors in the velocity field \citep{scharnowski2020}. Furthermore, significant Brownian motion can introduce uncertainty into PIV measurements \citep{olsen2007}. Another limitation is that tracer particles must be within an optimal range of density and size \citep{scharnowski2020}. This requirement can be impractical in biological systems using fluorescent proteins as markers, preventing the use of smaller window sizes as a workaround for issues related to Brownian motion or smaller turbulent flows \citep{kahler2012}. To overcome these limitations we explore a deep-learning-based optical flow (DLOF) algorithm for the estimation of the flow fields.   

In computer vision, optical flow describes the apparent motions of objects in a sequence of images \citep{barron1994}. Various rule-based techniques for optical flow estimation have been developed, including differential methods \citep{verri1990,bainbridge-smith1997,baraldi1996,bruhn2005a}, variational methods \citep{bruhn2005,cohen1993,bruhn2003,tu2019}, and feature-based methods \citep{cheng2002,cheng2006,beauchemin1995,becciu2009}. Specific implementations of those rule-based optical flow algorithms can be advantageous over PIV for applications in biological images \citep{bouguet2001,farneback2003,brox2004,zach2007,yong2021, vig2016a}. Rapid advancements in machine learning have resulted in deep learning optical flow (DLOF) algorithms, where the automatic feature extraction offered by deep convolutional neural networks has significantly improved the algorithm accuracy \citep{ranjan2017,fischer2015,ren2017,yu2023,bai2022,bar-haim2020,han2022,huang2022,hui2020,ilg2017,jeong2023,jiang2021,jiang2021a,liu2020,liu2021,luo2021,luo2022,min2023,nebisoy2021,pan2023,stone2021,sun2018,sun2022,ullah2019,xu2021,xu2022,zhao2020}.

Recent efforts used DLOF to estimate velocity fields in applications that would otherwise rely on PIV \citep{cai2019b,discetti2022,lagemann2021,yu2021,yu2023a,zhang2023}. In these instances,  DLOF was trained and evaluated using data from fluid dynamics simulations or computer-generated and augmented PIV datasets that mimic noisy data in real-world experiments. Obtaining ground-truth velocities required for training machine learning models is costly or impossible with real-world data. We investigate the performance of DLOF on experimental data of active nematic liquid crystals \citep{AditiSimha2002,Narayan2007,sanchez2012,guillamat2016,Kumar2018,Blanch-Mercader2018,Tan2019,Giomi2011,Giomi2012,Thampi2014,Shendruk2017}. We image microtubule (MT)-based active nematics under conditions that are beyond the limitations of PIV and present a significant challenge to its performance. We then develop a computational framework to apply DLOF to quantify the velocity fields of the MTs. We test the framework with ground truth velocity fields obtained by particle tracking methods. We compare the velocity fields obtained by PIV and DLOF against this ground truth data.

Microtubule (MT)-based active nematics are powered by ATP-consuming kinesin molecular motors. In such materials the extensile MT bundles generate internal active stresses, which in turn give rise to motile topological defects and associated autonomous flows\citep{sanchez2012}. Active nematics are described by two continuous fields, the director field, which describes the average orientation of the anisotropic MT filaments, and the velocity field, which describes their motions. Accurate measurement of the director field requires samples in which all the filaments are labeled. However, such samples yield low variations in spatial intensity, which makes application of PIV techniques challenging~\citep{tayar2022}. Specifically, in fully labeled active nematics PIV underestimates the velocity component parallel to the nematic director \citep{opathalage2019,memarian2024a,Tan2019,serra2023}. This can be attributed to the anisotropy of nematics; the intensity of MT bundles is fairly uniform along the nematic directors, which presents challenges for implementation of PIV. Alternatively, obtaining accurate PIV fields requires samples with a low volume fraction of labeled MTs, which creates highly speckled patterns suitable for PIV application, but from which the director field cannot be extracted. Overcoming these competing challenges requires active nematics containing high-concentration MTs in one color and dilute tracer MTs in a different wavelength~\citep{serra2023}. The former are suitable for director field measurement while the latter allow for accurate application of PIV techniques. However, these samples are cumbersome to prepare, and sequential imaging introduces a time lag between the measurement of the two fields.

We show that DLOF produces an accurate measurement of the flow field irrespective of the fraction of labeled filaments. Thus,  DLOF techniques can fully characterize the instantaneous state of an active nematic from one set of images. Furthermore, the  DLOF results are higher resolution and less noisy than those from PIV. These findings suggest that  DLOF models can be used for more accurate and robust measurements of the velocity field in diverse active and other soft matter systems.

\begin{figure*}
  \centering
  \includegraphics[width=\textwidth]{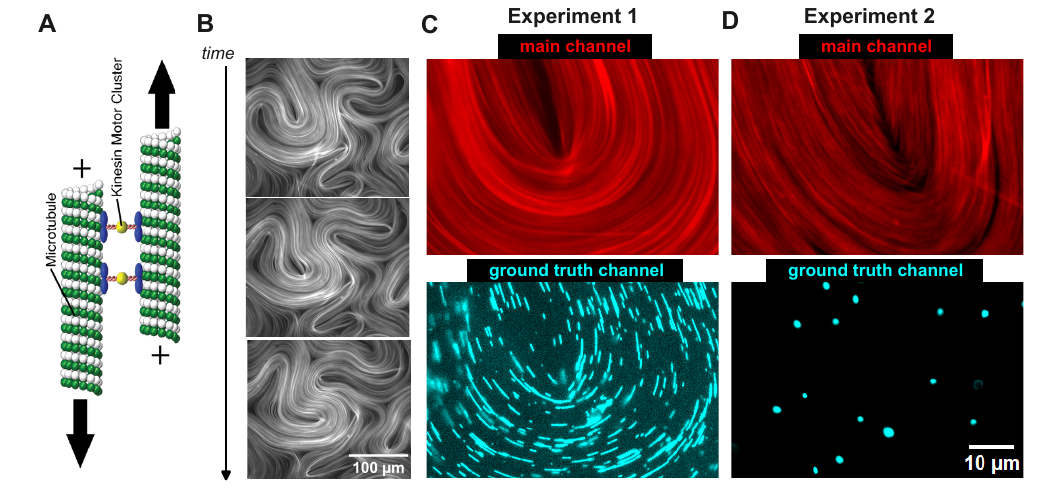}
  \caption{\textbf{Microtubule (MT)-based active nematics.}~(A) Microscopic components of the active nematic liquid crystal. Kinesin motor clusters consume energy to actively slide neighboring MTs against each other. (B) The active nematic exhibits the spontaneous flow that deforms the nematic texture over time. All MTs are fluorescently labeled at 647 nm. Increased local intensity indicates a higher local filament concentration. The time step is 7.5 s. (C) In Experiment 1, the fully labeled MTs (top panel) are mixed with a sparse population of MTs that fluoresce at 488 nm (bottom panel), which are used to generate ground-truth velocity points. (D) In Experiment 2, the fully labeled MTs (top panel) are mixed with passivated microbeads, which are used to generate the ground-truth velocities (bottom panel).}  
  \label{fig:nematics_system}
\end{figure*}

\section{Deep learning optical flow (DLOF)}


DLOF use convolutional neural networks for the automatic extraction of relevant features from the two adjacent frames in a video and use the extracted features to estimate the movements of objects between the two video frames \citep{ilg2017, ranjan2017,jonschkowski2020,yu2023, teed2020}. DLOF models are typically trained using supervised learning algorithms, in which training data are synthetic videos that include the true motions of all the objects in the videos across the video frames \citep{butler2012,menze2015,geiger2012,geiger2013,fritsch2013,dosovitskiy2015,mayer2016}. Synthetic data are required by this approach because obtaining the true displacements of objects in real-world videos is highly challenging. Thus, the ability of the models to properly adapt to unseen data from a different domain becomes crucial for the trained models to be useful in real-world scenarios. A recent study suggested that a model called RAFT (\textit{Recurrent All-pairs Field Transforms for Optical Flow}), which was originally trained using synthetic data, could generalize well to unseen fluid dynamics videos \citep{lagemann2021,teed2020}. However, this study evaluated the model's performance on simulation-generated videos and did not evaluate the performance on challenging videos obtained in experiments, such as the active nematics described above.

We evaluate the performance of RAFT on estimating velocity fields in active nematics experimental videos. In the sub-sections below, we briefly explain RAFT's architecture and relevant training methods. We describe experimental data collection in Section \ref{sec:experiments}. In section \ref{sec:results}, we present and discuss the benchmarks of RAFT against PIV on the active nematics experiments, using particle tracking data as ground-truth.

\subsection{Architecture of the RAFT model}

\begin{figure*} 
  \centering
  \includegraphics[width=\textwidth]{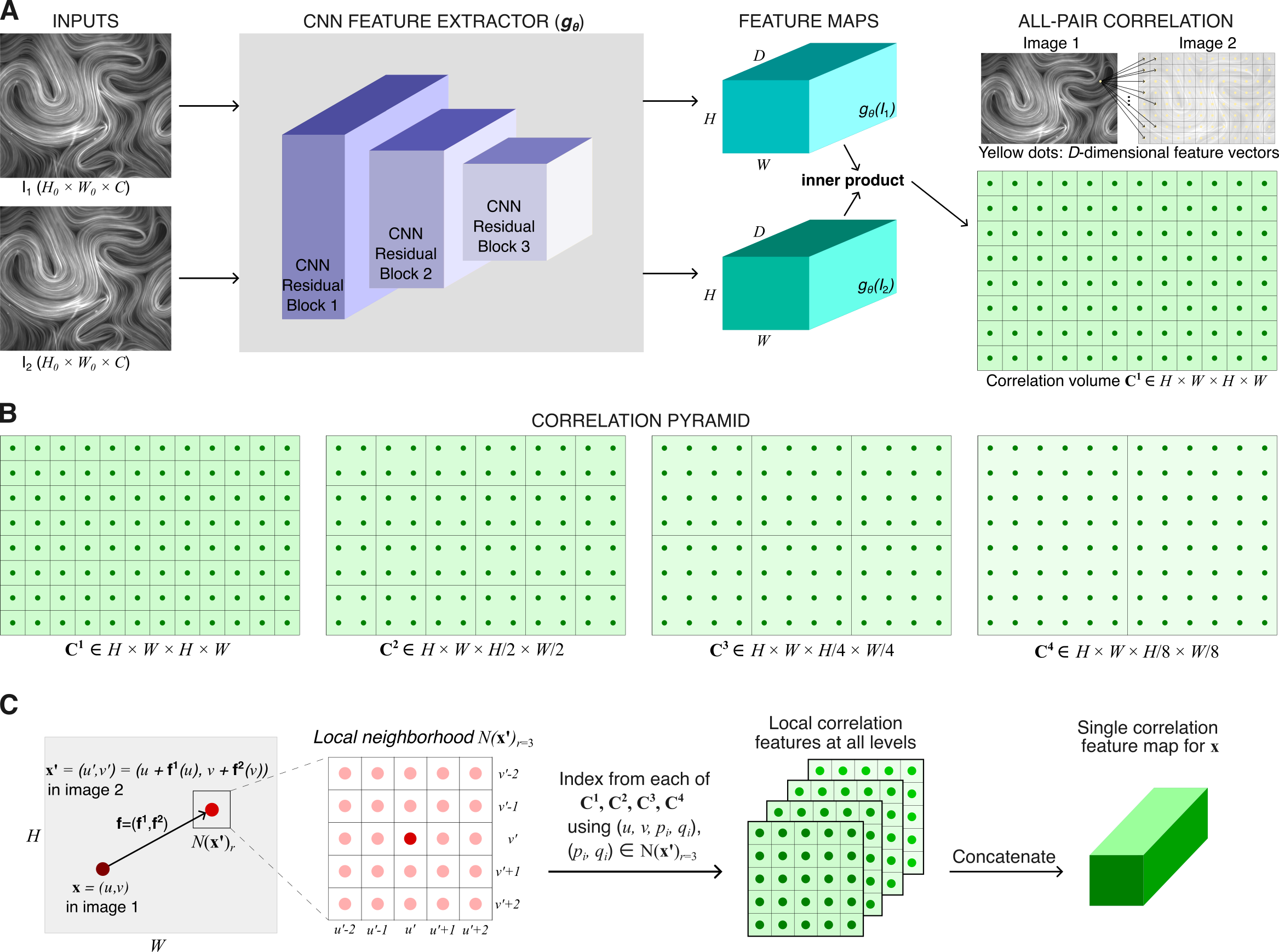}
  \caption{\textbf{Main components of the  DLOF model.}~(A) Feature extraction and construction of feature-level correlations: A convolutional neural network (CNN) is used to extract $D$ feature maps of resolution of $H \times W$ for each of the input images. Taking the inner product of the features maps of two images produces all-pair feature-level correlation volumes $\mathbf{C^1}$ of dimension $H \times W \times H \times W$. \textbf{(B) Correlation pyramid:} Multi-scale feature correlations are constructed by pooling the last two dimensions of $\mathbf{C^1}$, such that those dimensions are reduced by 1/2, 1/4, and 1/8, resulting in $\mathbf{C^2}$, $\mathbf{C^3}$, and $\mathbf{C^4}$, respectively. The first two dimensions preserve high-resolution information while multi-scale correlations enable the model to capture the motions of small fast-moving objects. \textbf{(C) Correlation lookup for a pixel $\mathbf{x}$ in $I_1$:} An estimate of the location of the correspondence $\mathbf{x'}$ (in $I_2$) is initialized by displacing $\mathbf{x}$ using the current flow estimate $\mathbf{f}$. The model then looks for the most correlated features in a neighborhood $\mathcal{N}(\mathbf{x'})_r$ centered at $\mathbf{x'}$ ($r=3$ in the figure), where all locations within $\mathcal{N}(\mathbf{x'})_r$ are used to index from the correlation pyramid $\{\mathbf{C^1}, \mathbf{C^2}, \mathbf{C^3}, \mathbf{C^4}\}$ to produce correlation features at all levels, which are further concatenated to form a single correlation feature map for the pixel $\mathbf{x}$ in $I_1$.} 
  \label{fig:model}
\end{figure*}

RAFT estimates the optical flow from a pair of images $(I_1, I_2)$ in three main stages: (1) Extract features of the input images using a convolutional neural network, (2) Use those extracted features to construct a correlation volume that computes the visual similarity of the images, and (3) Compute the final flow through an iterative process.

\textbf{\textit{Feature map extraction.}} 
The model uses an encoder $g_{\theta}$, which is a convolutional neural network, to extract features from the two input images. In particular, $g_{\theta}$ extracts features at 1/8 resolution; i.e., $g_{\theta}: \mathbb{R}^{H_0 \times W_0 \times C} \mapsto \mathbb{R}^{H_0/8 \times W_0/8 \times D}$, where $H_0$ and $W_0$ are the height and width of the images, $C$ the number of color channels ($C=3$ for RGB and $C=1$ for grayscale images), and $D$ the number of desired feature maps to be extracted. The encoder gradually reduces the resolution of the output feature maps; i.e., it successively outputs feature maps at 1/2, 1/4, and finally 1/8 resolution. For each of these steps, the resolution reduction is performed by convolutional residual neural network blocks (Figure \ref{fig:model}A). In general, feature maps produced at lower resolutions extract spatial correlations at higher levels with a wider receptive field, and it has been shown empirically that learning features at the aforementioned resolutions offers a balance between the model's performance and complexity \citep{teed2020}. 

\textbf{\textit{Construction of correlation feature map.}} Visual similarity between the two input frames is required to find the correspondences of moving objects between them. RAFT computes the visual similarity by constructing a correlation between all pairs of extracted features of first the image $g_{\theta}(I_1) \in \mathbb{R}^{H \times W \times D}$, and then that of the second image $g_{\theta}(I_2) \in \mathbb{R}^{H \times W \times D}$ (right part of Fig. \ref{fig:model}A). The elements of a correlation volume $\mathbf{C}(g_{\theta}(I_1), g_{\theta}(I_2) \in \mathbb{R}^{H \times W \times H \times W}$ are given by $C_{ijkl} = \sum_{h} g_{\theta}(I_1)_{ijh} \cdot g_{\theta}(I_2)_{klh}$. Correlations are further computed as a 4-layer pyramid $\{\mathbf{C^1}, \mathbf{C^2}, \mathbf{C^3}, \mathbf{C^4}\}$, where $\mathbf{C}^k$ has dimensions $H \times W \times H/2^k \times W/2^k$ (Figure \ref{fig:model}B). Here, the reduction of the last two dimensions of the correlation volume $\mathbf{C}$ by a factor of $2^k$ is achieved by pooling the last two dimensions of $\mathbf{C}$ with kernel size $k$ and equivalent stride. Having correlations at multiple levels through $\{\mathbf{C^1}, \mathbf{C^2}, \mathbf{C^3}, \mathbf{C^4}\}$ allows the model to handle both small and large displacements. The first two dimensions (that belong to $I_1$) are maintained to preserve high-resolution information, enabling the model to detect motions of small fast-moving objects.  

The link between an object in $I_1$ and its estimated correspondence in $I_2$ is determined through correlation lookup using the correlation pyramid, as described in Fig. \ref{fig:model}C. 
The correspondence $\mathbf{x'} \in I_2$ of a pixel $\mathbf{x} = (u,v) \in I_1$ is estimated by $\mathbf{x'} = (u + f^1(u), v + f^2(v))$, where $(\mathbf{f^1}, \mathbf{f^2})$ is the current estimate of  DLOF  between $I_1$ and $I_2$. 
A local grid around $\mathbf{x'}$ is then defined as $\mathcal{N}(x')_r = \{ \mathbf{x'} + \mathbf{dx}\ \vert \ \mathbf{dx} \in \mathbb{Z}, \|\mathbf{dx}\|_1 \leq r \}$, a set of integer offsets that are within a radius of $r$ of $\mathbf{x'}$ (using $\ell_1$ distance). 
The local neighborhood $\mathcal{N}(\mathbf{x'})_r$ is used to index from all levels of the correlation pyramid using bilinear sampling, such that the grid $\mathcal{N}(\mathbf{x'}/2^k)_r$ is used to index the correlations $\mathbf{C}^k$. 
At a constant searching radius $r$ across all levels, a local neighborhood on a lower level implies a larger context; for example, at $k=4$, a neighborhood of $r=4$ effectively includes a range of 256 pixels at the video's resolution.   
The interpolated correlation scores at all levels are concatenated to form a single feature map, which serves as an input for iterative flow refinement described below.

\textbf{\textit{Iterative flow refinement.}}
The flow between the two input images is determined through an iterative process, such that the final flow $\mathbf{f}_N$ is obtained from the sequence $\mathbf{f}_{k+1}=\mathbf{f}_k +  \Delta \mathbf{f}$ where $0 \leq k \leq N-1$, $\mathbf{f}_0 = \mathbf{0}$, $N$ is the number of iterations, and $\Delta \mathbf{f}$ is being produced by the model at each of the iterations. The flow updating is performed by a convolutional Gated Recurrent Unit (ConvGRU) cell \citep{dey2017}, in which convolutions have replaced fully connected layers:

\begin{equation}
    z_t = \sigma( \textrm{Conv}_{3 \times 3}([h_{t-1}, x_t], W_z) )
\end{equation}
\begin{equation}
    r_t = \sigma( \textrm{Conv}_{3 \times 3}([h_{t-1}, x_t], W_r) )
\end{equation}
\begin{equation}
    \tilde{h}_t = \textrm{tanh}(\textrm{Conv}_{3 \times 3}([r_t \odot h_{t-1}, x_t], W_h))
\end{equation}
\begin{equation}
    h_t = (1 - z_t) \odot h_{t-1} + z_t \odot \tilde{h}_t
\end{equation}
where $x_t, z_t, r_t, \tilde{h}_t, h_t$ are the input, update gate, reset gate, internal memory state, and hidden state at time $t$, respectively; $\sigma(\cdot)$ is the sigmoid function, $\textrm{tanh}(\cdot)$ the hyperbolic tangent, and $\textrm{Conv}_{3\times3}(\cdot, W)$ the convolution operator with kernel size $3\times3$ and bias $W$. Here, the hidden state $h_t$ is further processed by two convolutions to produce the flow update $\Delta \mathbf{f}$ at time $t$. 

In the above set of equations, at a current time $t$, the input $x_t$ is the concatenation of the current flow estimate, correlation, and context features. 
The update gate $z_t$, which is calculated using the last hidden state $h_{t-1}$ and the current input $x_t$, controls how much past knowledge should be considered in the computation of the current hidden state $h_t$. 
The reset signal $r_t$ is a function of the current input $x_t$ and the last hidden state $h_{t-1}$, and determines how much of the past knowledge to forget.
The internal memory $\tilde{h}_t$ of the GRU cell is calculated using the current input $x_t$ and the last hidden state $h_{t-1}$ weighted by the reset gate $r_t$. 
Finally, the hidden state is updated by the weighted sum of the last hidden state $h_{t-1}$ and the current cell memory $\tilde{h}_t$, with the update gate $z_t$ controling the weights distribution.

\subsection{Training DLOF}
Most state-of-the-art DLOF models are trained via supervised learning using synthetic data, where flow ground truths can be obtained straightforwardly during data generation. The supervised loss $\mathcal{L}_\text{s}$ used to optimize RAFT's parameters compares the sequence of predictions $\{ \mathbf{f}_1, ..., \mathbf{f}_N \}$ with the flow ground truth $\mathbf{f}_\text{gt}$, with exponentially increasing weights:
\begin{equation}
    \mathcal{L}_\text{s} = \sum_{i=1}^{N} \gamma^{N-1} \| \mathbf{f}_\text{gt} - \mathbf{f}_i\|_1
\end{equation}
where $\gamma < 1$.

RAFT is trained using supervised learning, and it has been shown to generalize well to data in other domains \citep{teed2020,lagemann2021}. When it is required, the model's parameters can be further fine-tuned using the real-world data in the target domain; however, unsupervised learning is generally required because ground truths of those data are often unavailable. 

\textit{Unsupervised training.} An approach to unsupervised training is to generate realistic pseudo-flow ground truth data using the current model, and then use that pseudo ground truth data for further optimizing the model's parameters. In this approach, the current model is first used to warp the image $I_1$ to produce an estimate of the image $I_2$, i.e., $\tilde{I}_2 = \Omega(I_1, \mathbf{f})$ where $\Omega$ is the warping function that displaces the pixels in $I_1$ according to the current estimate  $\mathbf{f}$ of the flow. $\tilde{I}_2$ can be then used as a pseudo ground truth to compute a simple unsupervised loss 
\begin{equation}
    \mathcal{L}_\text{u} = w_\text{photo} \cdot \mathcal{L}_\text{photo} + w_\text{smooth} \cdot \mathcal{L}_\text{smooth}
\end{equation}
where $\mathcal{L}_\text{photo}$ denotes the photometric loss between $I_2$ and $\tilde{I}_2$, $\mathcal{L}_\text{smooth}$ flow smoothness regularization, and $w_\text{photo}$, $w_\text{smooth}$ are the weights. The photometric loss quantifies the structural and visual differences between $I_2$ and $\tilde{I}_2$, being aware of occluded regions in which pixels in $I_1$ do not have their correspondences in $I_2$. A common metric used for photometric loss is the occlusion-aware structural similarity index (SSIM) \citep{gordon2019,brox2004}. A major challenge in unsupervised training of  DLOF models is to obtain an accurate estimate of occlusions \citep{jonschkowski2020}, which cannot be directly measured when dealing with real-world data. The unsupervised loss above also has a second term to encourage the smoothness of the resultant velocity fields. For example, the $k$th order smoothness is defined as \citep{jonschkowski2020} 
\begin{equation}
    \mathcal{L}_{\text{smooth}(k)} = \frac{1}{n} \sum \textrm{exp}\left( - \dfrac{\nabla I}{\sigma}  \right) \cdot \| \nabla^{(n)} V \|
\end{equation}
where $\nabla I$ detects the edges in the current image, $\nabla^{(k)} V$ is the $k$-th order gradient of the corresponding velocity field, $\sigma$ controls the strength of the regularization, and $n$ is total number of samples.

We obtained the results in the benchmarks of this work using a RAFT model that was trained with the FlyingThings synthetic datasets \citep{mayer2016}, which yielded the highest performance in our investigation. During velocity computation, we empirically set the number of iterations for flow refinement to 24.

\section{Active nematics samples }
\label{sec:experiments}

We tested the performance of the  DLOF framework using a MT-based active nematic liquid crystal~\citep{sanchez2012}. An active nematic is a quasi-2D liquid crystal comprised of locally aligned filamentous MTs. When powered by kinesin molecular motors, extensile MTs spontaneously generate a chaotic flow field that varies over space and time, and in turn reorients the nematic texture (Fig. 1A-B). Typically, the velocity field is computed by performing PIV on images of active nematics comprised of fluorescently labeled MTs. However, this method can be inaccurate when all the MTs are labeled, as these samples have poor contrast variations in fluorescence intensity, especially in the direction of the MT alignment~\citep{tayar2022}.

We performed two distinct experiments, each containing a different type of tracer that we used to estimate the ground truth. In both experiments, a large fraction of MTs were labeled with a fluorescent dye that emits 647 nm wavelength photons. In Experiment 1, samples contained a very low concentration of 488 nm labeled MTs. They were dilute enough so that individual filaments could be distinguished (Fig. 1C, bottom panel). However, accurately linking the detected MTs into time trajectories was only possible for a small fraction of the dilute population. In Experiment 2, instead of relying on dilute labeling, we mixed passivated 488 nm fluorescent microbeads into the active nematic (Fig. 1D, bottom panel). Although not directly incorporated into the quasi-2D active nematic, these beads were located right above the nematic layer and followed the same flow field. Compared to the sparsely-labeled MTs, the beads could be reliably tracked across several frames with an automated algorithm [SI], thus providing a larger set of velocity values that served as the ground truth. 

\section{Results \& Discussion}
\label{sec:results}

\subsection{Experiment I: ground truth provided by sparsely labeled MTs}

\begin{figure*}
  \centering
  \includegraphics[width=\textwidth]{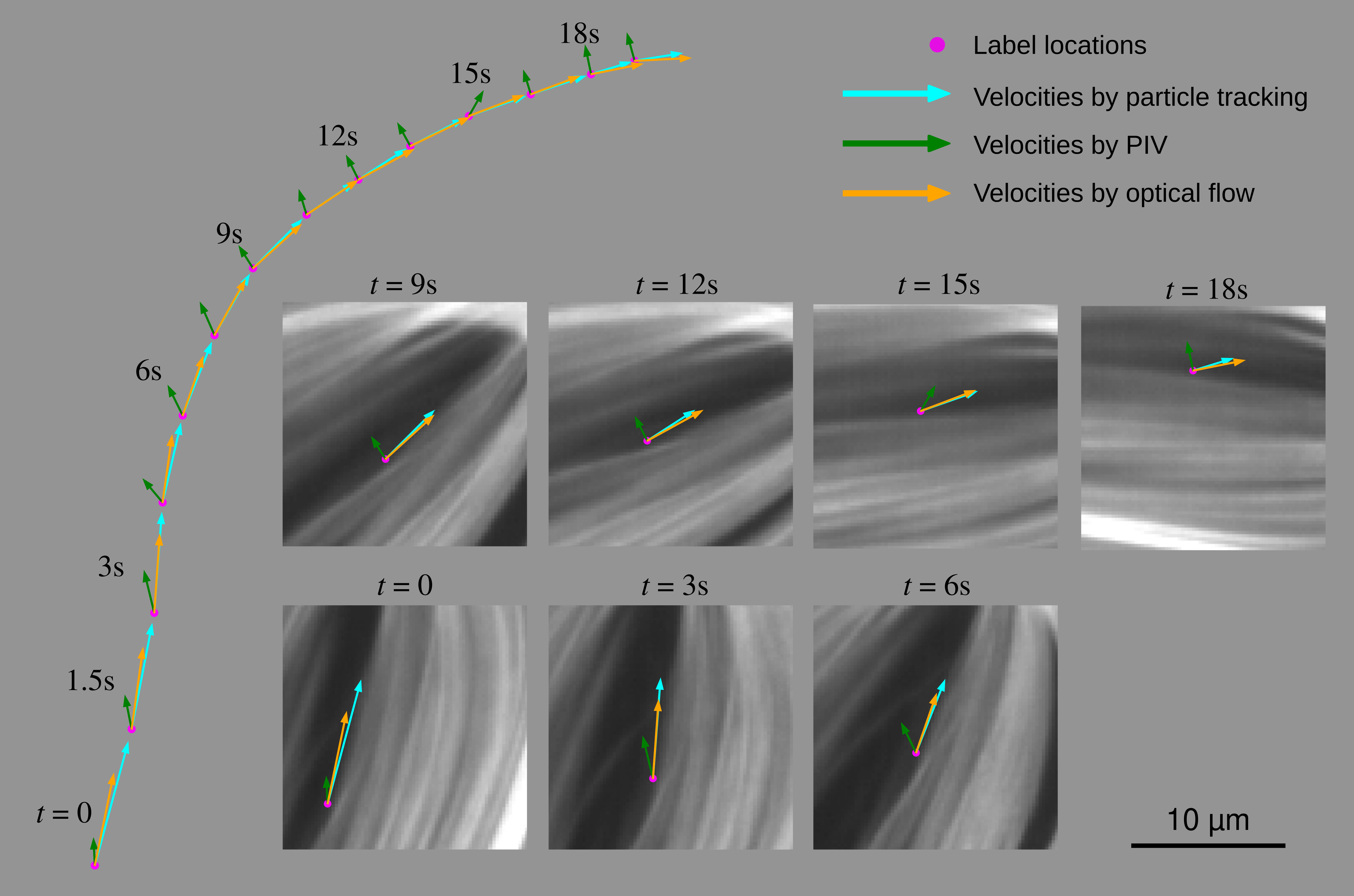}
  \caption{\textbf{DLOF outperforms PIV for densely labeled samples.}~(left) The trajectory of an individual MT, which is imaged every 1.5 seconds. MT true velocities (cyan arrows) are obtained by particle tracking. The velocity vectors estimated by PIV and DLOFs are indicated respectively with green and orange arrows. The insets depict the densely labeled MTs in local neighborhoods of the tracked labels at the indicated times. The high densities of the labels in the images pose a significant challenge to PIV, resulting in inaccurate velocity estimates. In contrast, DLOF produces highly accurate velocities. Particle tracking was extracted from a simultaneously imaged sparsely labeled channel. 
  }\label{fig:example_trajectories}  
\end{figure*}

We first studied active nematics containing both densely and sparsely labeled MTs with different fluorophores. The sample was imaged sequentially in the dense and sparse channel. Using these samples we first performed PIV and DLOF on densely labeled samples. This data was compared to particle tracking of sparsely-labeled active nematics, which served as ground truth (Fig. \ref{fig:example_trajectories}). We found that the velocities estimated by PIV for densely labeled systems are highly inaccurate. DLOF overcomes this limitation of PIV and provides more accurate estimates of both the magnitude and the direction of the velocity. We hypothesize that the breakdown of the PIV for densely labeled systems arises because the algorithm cannot reliably distinguish contrast variations at high densities. As we show below, the breakdown is strongest in directions parallel to the director field. PIV significantly underestimates the velocity tangent to the MT bundles because the contrast is more uniform in that direction, as was previously reported \citep{opathalage2019,memarian2024a,Tan2019,tayar2022,serra2023}. 

\begin{figure*} 
  \centering
  \includegraphics[width=\textwidth]{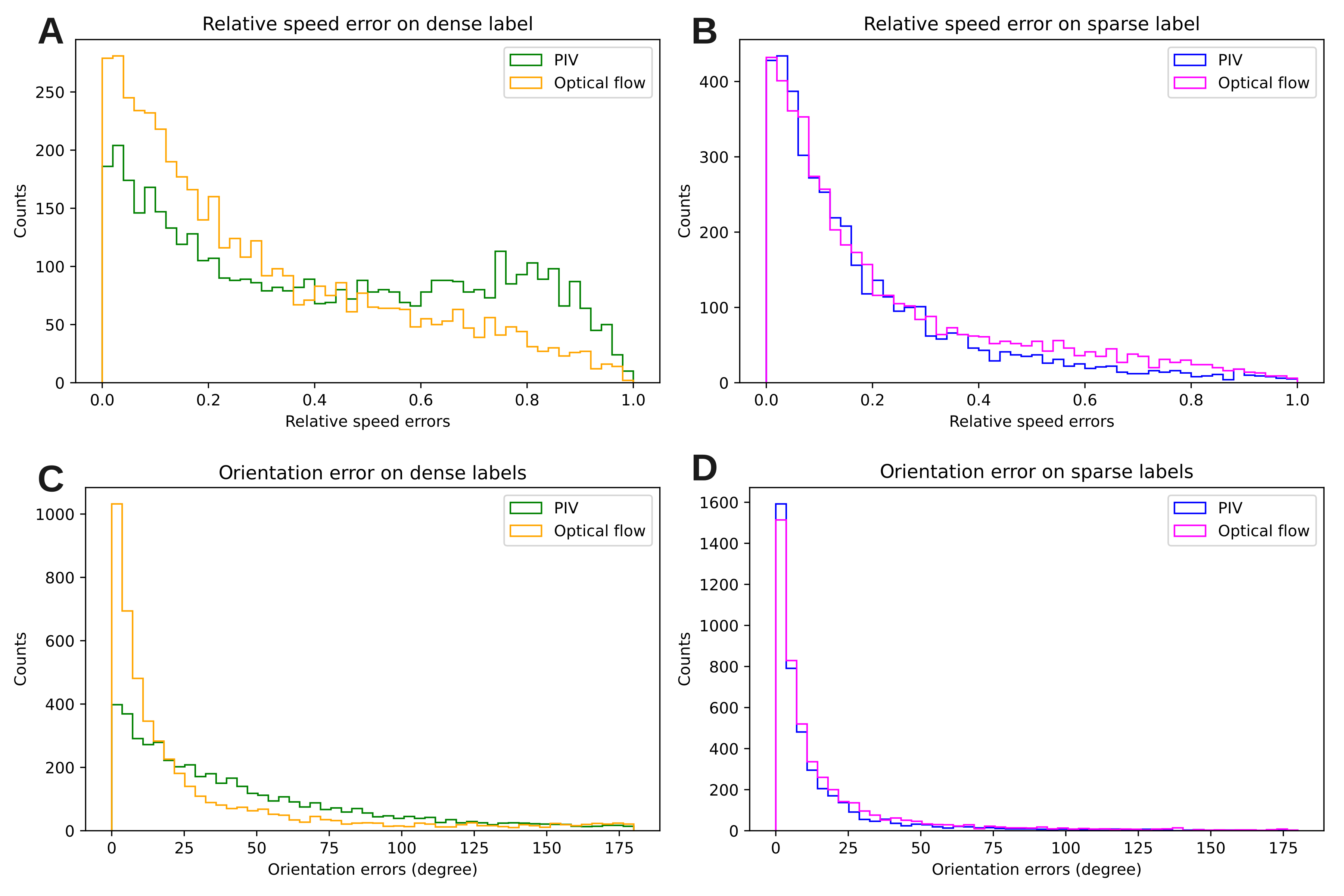}
  \caption{\textbf{Comparing PIV and DLOF to single-filament tracking.}~Distribution of errors when comparing PIV and DLOF velocity fields from sparsely and densely labeled samples to single-filament tracking. The distributions of errors in the magnitude and orientation of the velocity (defined in the text) for PIV and DLOF. Errors are computed by comparing different estimates with particle tracking results. The mean relative speed errors for PIV are 42\%  and 19\% for densely and sparsely labeled systems; errors for DLOF are 29\% and 23\%. The mean orientation errors for PIV are 44 degrees and 14 degrees for densely and sparsely labeled systems; errors for DLOF are 29 degrees and 17 degrees. The distributions are obtained from 4738 traced labels across 44 frames in Experiment 1.}
  \label{fig:error_hists}
\end{figure*}

To quantify the above-described observations, we used PIV and DLOF to estimate the velocity fields from the dense and dilute channels. We compared these to the ground truth based on single-particle tracking. PIV and DLOF estimate the flow field everywhere while single particle tracking yields velocities only at the location of tracked points. For each of the inter-frame displacements of the traced labels, the velocity magnitude error is calculated by $|||\mathbf{v}|| - ||\mathbf{v}^*||| / ||\mathbf{v}^*||$ where $\mathbf{v}^*$ is the true displacement vector obtained from particle tracking at a particular position and $\mathbf{v}$ is the velocity obtained at the same position from either the PIV or DLOF. The orientation error $\theta$ is calculated using the cosine similarity, where $\cos(\theta) = \mathbf{v} \cdot \mathbf{v^*} / (||\mathbf{v}|| \cdot ||\mathbf{v^*}||)$. By repeating the procedure for all tracked particles we obtained the distribution of measurement errors (Figs.~\ref{fig:error_hists}). PIV and DLOF have comparable errors for sparse labels (Fig.~\ref{fig:error_hists}B). Specifically, using the sparse labels, PIV and  DLOF have mean relative speed errors of $19\%$ and $23\%$, respectively. 
However, with dense labels, PIV results were more unreliable, with relative speed errors extending out to $100\%$, and the mean relative speed error increased to $42\%$.  
In contrast, the DLOF estimates are nearly as good as those with sparse labels, with a mean relative speed error of $29\%$ (Fig.~\ref{fig:error_hists}A).  
Similarly, the mean orientation errors of PIV and DLOF are also comparable when using sparse labels, $14$ and $17$ degrees, respectively (Fig.~\ref{fig:error_hists}D). The discrepancy between orientation errors produced by PIV and DLOF becomes significant when using dense labels, where the mean orientation error of PIV increases to $44$ degrees while that of DLOF is only $29$ degrees (Fig.~\ref{fig:error_hists}C).

\begin{figure*} 
  \centering
  \includegraphics[width=0.95\textwidth]{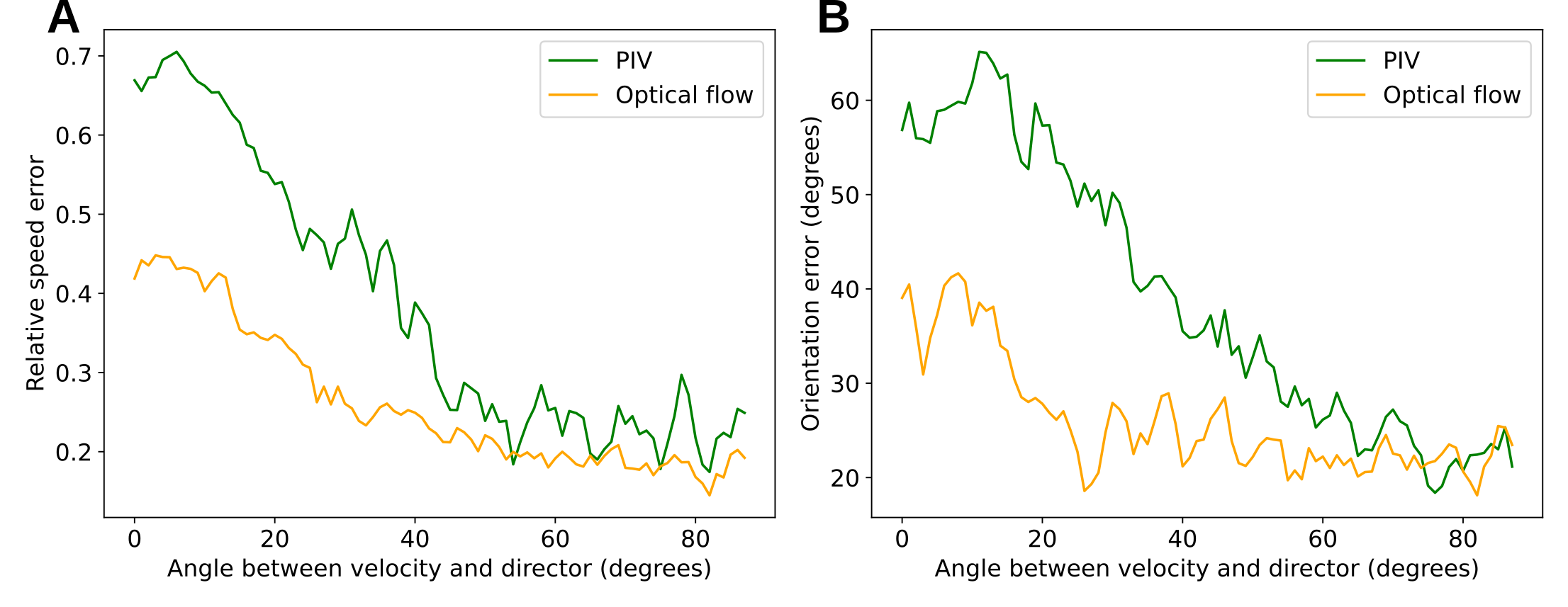}
  \caption{\textbf{The improvement of DLOF over PIV increases as the velocity becomes parallel to the director field (for dense labels).}~Average relative speed error (A) and average orientation error (B) of PIV and DLOF as a function of the angle between ground truth velocity and director. PIV particularly breaks when the velocities are tangent to the MT bundles due to the uniform contrast of the dense labels along MT bundles. DLOF can handle the uniform contrast along MT bundles and thus produces much more accurate velocities.
  }
  \label{fig:error_velocity_alignment}
\end{figure*}

Previous studies \citep{opathalage2019,memarian2024a,Tan2019,serra2023} had shown that uniform contrast along densely labeled MT bundles poses a major challenge to PIV, resulting in significantly underestimated velocity component tangent to the MT bundles. We therefore evaluated the contribution of this effect to our observed breakdown of PIV as follows. We extracted the director, i.e., the local orientation, of the MT bundles using the dense labels and computed average errors of velocities obtained by PIV and DLOF as functions of the angle between ground truth velocity and director (Fig. \ref{fig:error_velocity_alignment}). We find that when the MTs are moving in directions with significant components along the director, PIV produces high relative speed errors (Fig. \ref{fig:error_velocity_alignment}A) and orientation errors (Fig. \ref{fig:error_velocity_alignment}B). DLOF strongly improves the estimation of velocities in these directions. In particular, the improvement of DLOF over PIV uniformly increases as the velocity direction approaches the director field. When the velocities are parallel to the directors (i.e., angles between velocity and director are less than 1 degree), the average relative speed error is reduced by 37\% with DLOF (compared to PIV), and average orientation error reduced by 31\%. This analysis shows that DLOF can resolve this well-known limitation of PIV, and thus establishes DLOF as an alternative method capable of obtaining accurate velocity fields with dense labels.

\paragraph*{Comparing PIV and DLOF spatial flow fields.}

\begin{figure*} 
  \centering
  \includegraphics[width=\textwidth]{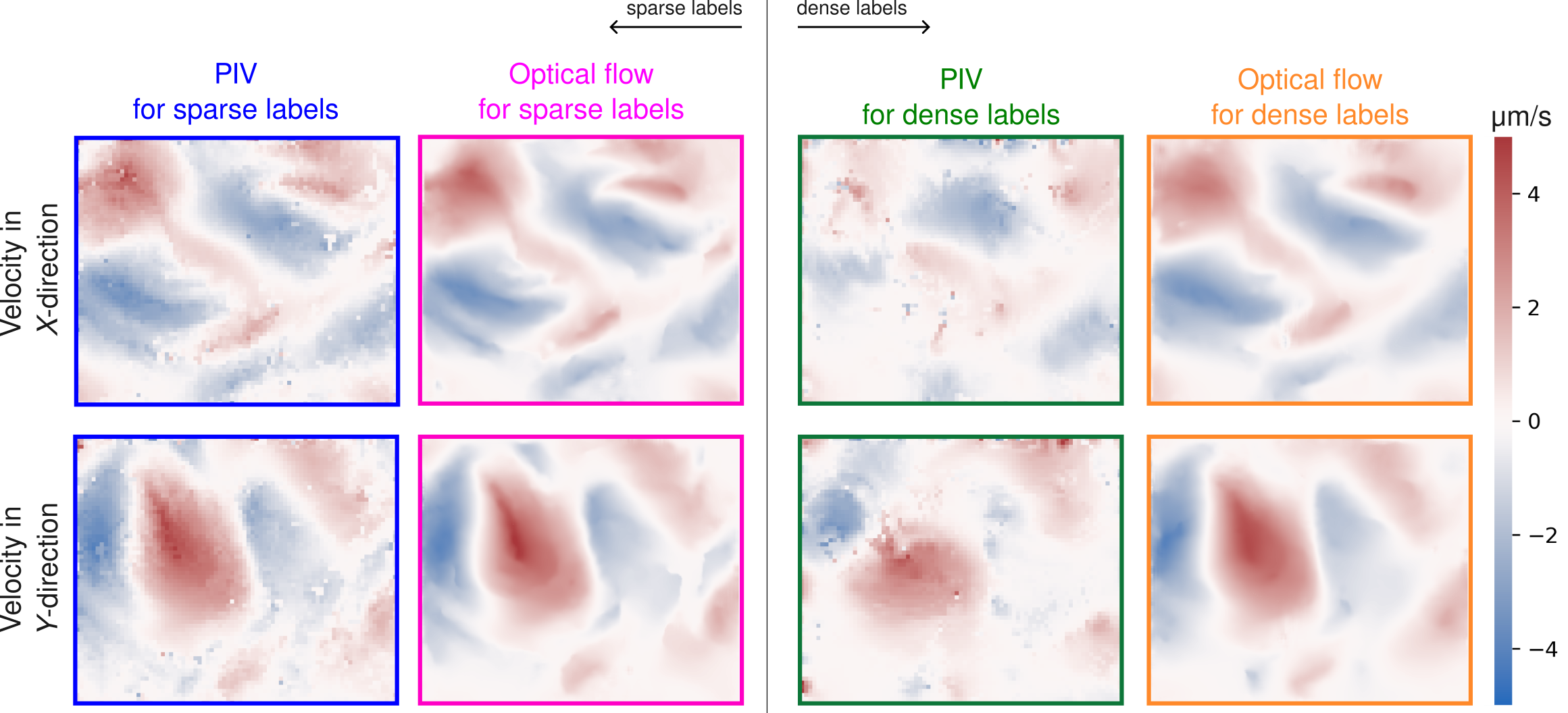}
  \caption{\textbf{Comparison of the velocity fields in the x-direction (top row) and y-direction (bottom row) produced by PIV and  DLOF for sparse labels (blue and magenta highlighted), and by PIV and  DLOF for dense labels (green and orange highlighted).}~The velocity fields are calculated for the first frame obtained from Experiment 1.  DLOF always produces smoother fields, due to its capability to estimate displacements on a pixel-level. Remarkably, when dealing with dense labels, velocity fields estimated by  DLOF are significantly more accurate than those produced by PIV (by comparing green and orange boxes for each velocity component).}
  \label{fig:qualitative_comparison}
\end{figure*}

Thus far, our analysis has focused on the accuracy of the PIV and DLOF methods in estimating the velocities of individual traced labels. Next, we evaluate the quality of the two-dimensional flow fields produced by each method. In this case, we do not have ground truth to compare against, since the tracked dilute MTs do not yield the spatial flow fields. The analysis described above showed that PIV and DLOF are comparable for sparsely labeled systems. Therefore, we use the flow fields determined by PIV with sparse labels as the baseline. For a meaningful comparison, we note that PIV produced the velocity fields on lower-resolution spatial grids when compared to DLOF. Therefore, we interpolate the DLOF results onto the lower-resolution grid of the PIV results. PIV and DLOF produce consistent flow fields for sparsely labeled samples (Fig~\ref{fig:qualitative_comparison}). However, the DLOF results are significantly smoother. While the DLOF results are somewhat noisier for the densely labeled system, the correct flow structure is maintained. In comparison, PIV on densely labeled systems produces an inaccurate flow structure. Importantly, the  DLOF model correctly estimates velocities across different regions and different scales of the flow speed. For example, MT bundles move faster in the vicinity of $+1/2$ topological defects and slower near $-1/2$ defects.

\begin{figure} 
  \centering
  \includegraphics[width=\columnwidth]{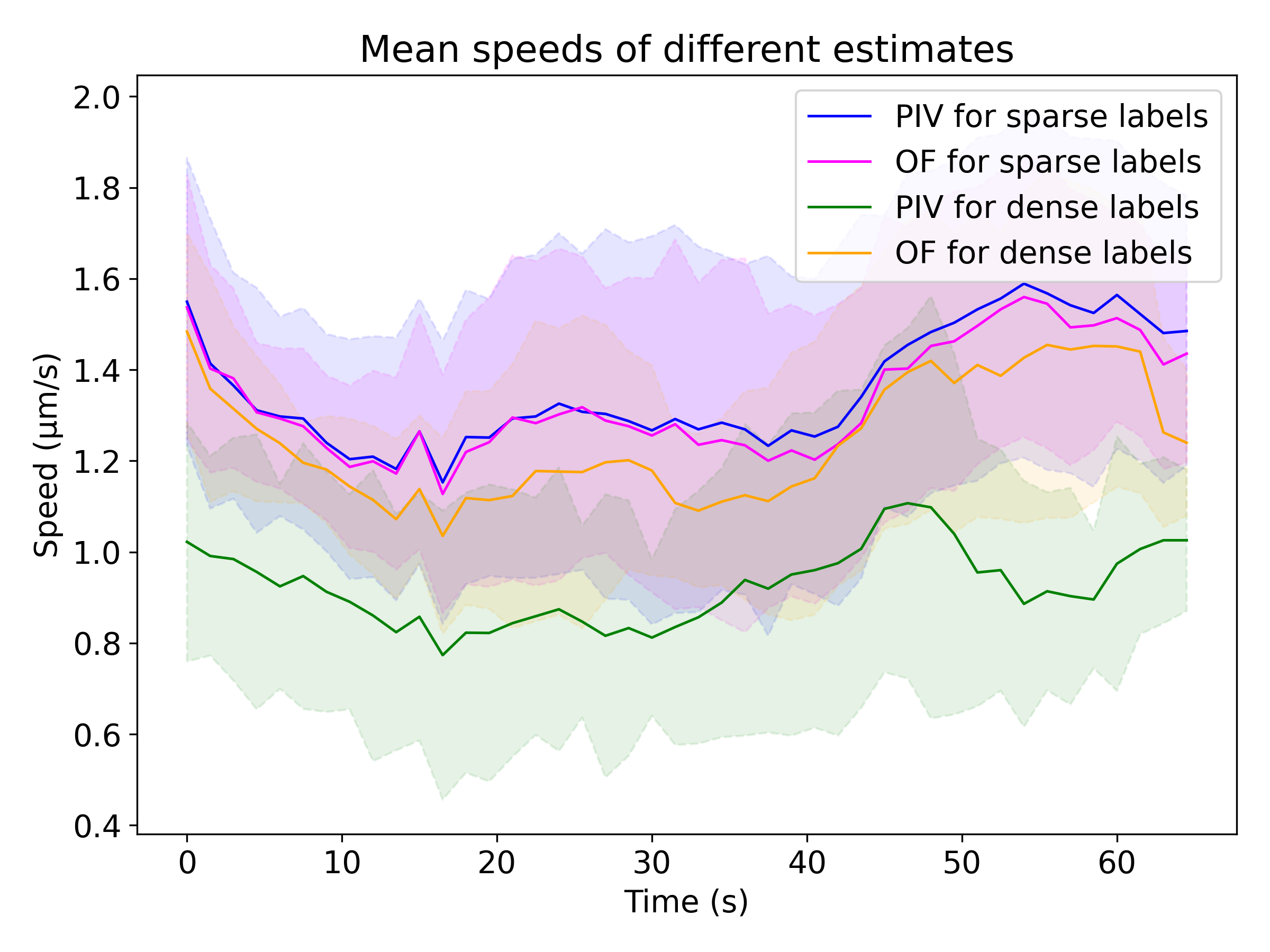}
  \caption{\textbf{Comparison of mean flow speeds as a function of time.} The flow speeds ($\mu$m/s) averaged over the entire spatial domain are shown as a function of time over the 44 frames of the benchmark video using the dense labels. The frame interval is 1.5 seconds, and results are shown for sparse and dense labels for PIV and optical flow. The shaded areas show $95\%$ confidence levels of the mean speeds.}
  \label{fig:mean_flow_speed}
\end{figure}

\begin{figure} 
  \centering
  \includegraphics[width=\columnwidth]{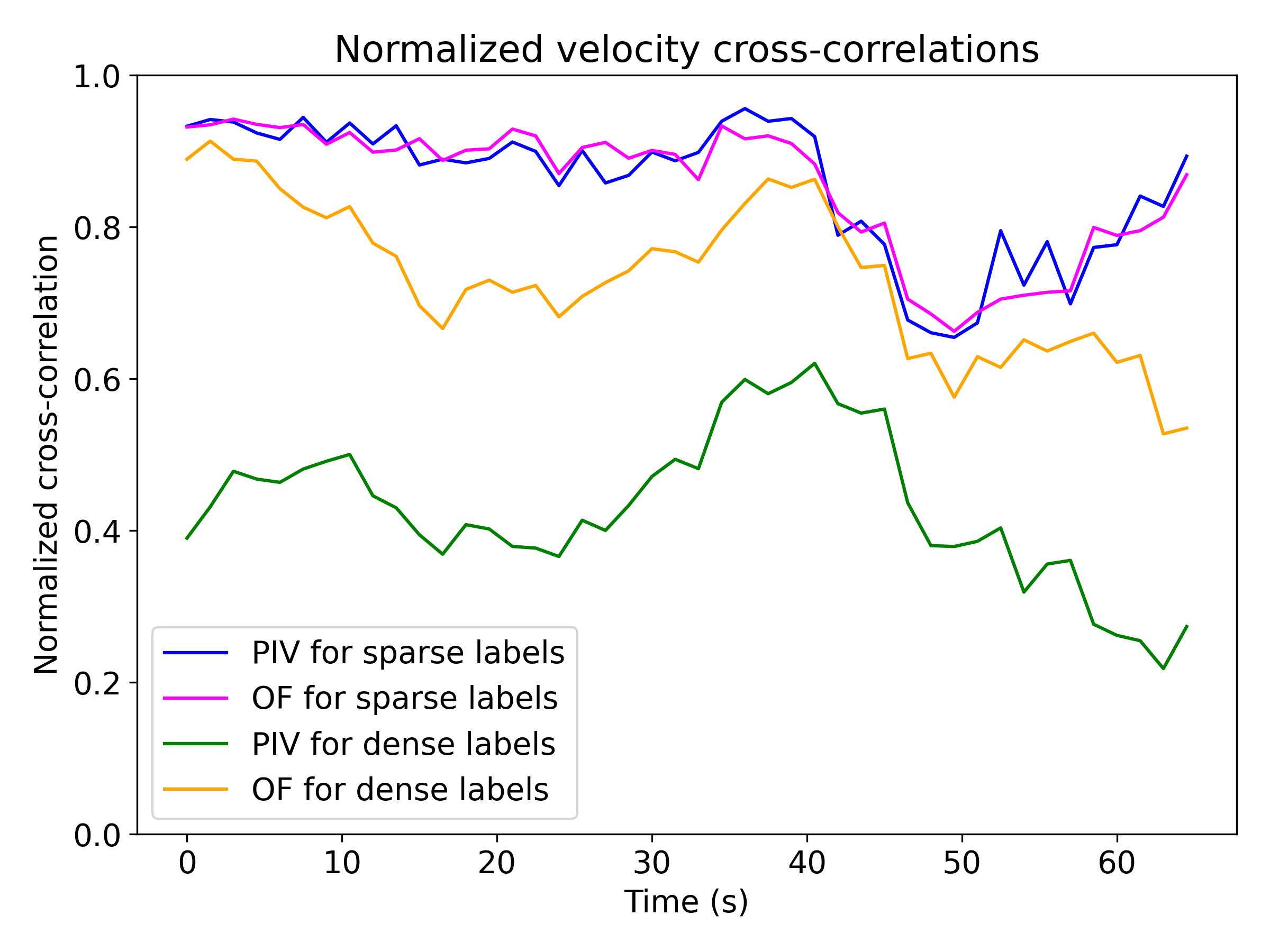}
  \caption{\textbf{Normalized zero-lag cross-correlation between velocity estimates and ground truth.}~The normalized spatial correlation (Eq.~\eqref{eq:crosscorr}) is shown for optical flow on sparsely and densely labeled systems, as well as PIV on densely labeled systems, as a function of time.}
  \label{fig:spatial_corr}
\end{figure}

We compared the flow speeds obtained from PIV and DLOF averaged over the entire field (Fig.~\ref{fig:mean_flow_speed}). Consistent with the previous analysis above, the PIV and DLOF estimates are nearly identical for sparsely labeled samples. The  DLOF estimates for dense labels fall within the 95\% confidence interval. In contrast, PIV significantly underestimates the velocities for dense labels.

As a further comparison between different methods, we define the normalized zero-lag cross-correlation between an estimated velocity and the ground truth as
\begin{align}
\mathcal{C} = \frac{\sum_{i} \mathbf{v}_i \cdot \mathbf{v}_i^*}  {\sum_{i} \mathbf{v}_i^* \cdot \mathbf{v}_i^*} = \frac{\sum_{i} \mathbf{v}_i \cdot \mathbf{v}_i^*}  {\sum_{i} \| \mathbf{v}_i^* \|^2},
\label{eq:crosscorr}
\end{align}
where $\mathbf{v}_i$ and $\mathbf{v}_i^*$ are the estimated and the ground truth velocities of the traced label $i$, and $\sum_{i}$ sums over all the traced labels in the current frame. A perfect velocity estimation would yield $\mathcal{C}=1$, while $\mathcal{C}>1$ indicates that, on average, flow speeds are overestimated and $\mathcal{C}<1$ underestimated. PIV and DLOF perform similarly for sparse labels (Fig. \ref{fig:spatial_corr}). The performance discrepancy between PIV and  DLOF becomes significant for dense labels, where velocities produced by  DLOF are still highly correlated with the ground truths. In contrast, velocities estimated by PIV result in significantly lower correlations.

\subsection{Experiment 2: ground truth provided by passive beads}
\begin{figure}
  \centering
  \includegraphics[width=\columnwidth]{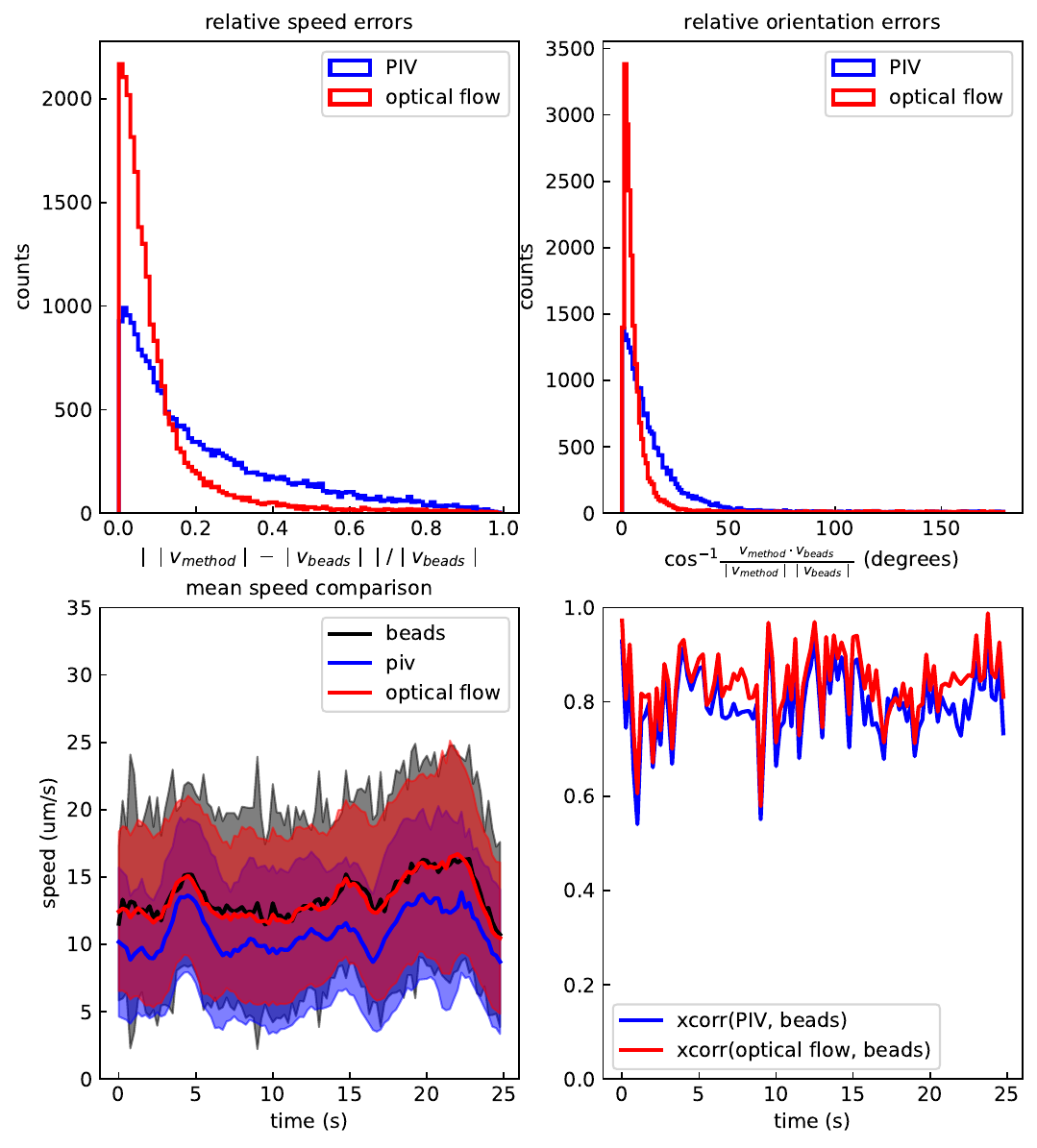}
  \caption{\textbf{Comparison of PIV and DLOF where passive tracer beads generate the ground-truth velocities.}~(A) Histograms of speed differences between PIV and bead velocities, and between optical flow and bead velocities. (B) Histograms of angular orientation differences between PIV and bead velocities, and between optical flow and bead velocities. (C) Mean speed of the beads, PIV, and optical flow over time. The speed is averaged over all available points for the given field (note that there are far more optical flow points than PIV points, and far more PIV points than beads, in each frame). Error bars indicate the standard deviation. 
  (D) Zero-lag cross correlation between PIV and bead velocities, and between PIV and optical flow  velocities over time (Eq.~\eqref{eq:crosscorr}).}
  \label{fig:bead_tracking_results}
\end{figure}

We also compared DLOF and PIV against tracked passive beads, which served as the ground truth. In each frame, we compared the instantaneous velocity of each bead to the velocities at the same position generated by PIV and DLOF. Since we computed PIV on a sparse grid, we interpolated its values as necessary to correspond to bead positions. As in Experiment 1, the comparison shows that DLOF is more accurate than PIV (Fig. \ref{fig:bead_tracking_results}). In particular, the difference in speeds between the beads and the  DLOF velocities was significantly smaller than that between the beads and PIV (Fig. \ref{fig:bead_tracking_results}A). Similarly, the angular orientations of DLOF velocities were also closer to the bead velocities (Fig. \ref{fig:bead_tracking_results}B). At each time point, the spatially averaged mean speed of the DLOF field was closer to that of the beads, while the mean speed of PIV was systematically lower (Fig. \ref{fig:bead_tracking_results}C). This result is consistent with the notion that PIV systematically underestimates the motion of MTs when their motion is locally parallel, rather than perpendicular, to intensity gradients in the image on a length scale larger than the size of PIV's interrogation region \citep{scharnowski2020,raffel2018}.  Lastly, the zero-lag cross-correlation  Eq.~\eqref{eq:crosscorr} between the DLOF and bead velocities was consistently higher than the correlation between PIV and bead velocities (Fig. \ref{fig:bead_tracking_results}D).

Our benchmarks demonstrate the accuracy of DLOF for extracting velocities from active nematics, surpassing the limitations of traditional PIV methods. Although we have trained and demonstrated the model on 2D active nematics samples captured with a $60\times$ magnification objective, we note that it appears to generalize well to other magnifications and situations, such as 2D slices from a 3D isotropic active MT system \cite{sanchez2012} captured at lower magnification ($10 \times$), provided that: there is sufficient contrast between labeled MTs and the background, the illumination of MTs does not change significantly between the two input frames, and the movements between two input frames are smaller than the algorithm's search window and the scale of the moving textures in the images.

\section{Conclusions}
We compared DLOF and PIV for estimating the velocity fields of active nematics. DLOF produces spatially smoother velocity fields. It also generates more accurate flows than PIV for high densities of fluorescent filaments. Furthermore, unlike PIV, DLOF eliminates the need to manually tune and readjust the model's parameters when working with data that have high contrast variances across the entire data. This is essential for analyzing large amounts of data, or for real-time control applications where it is impractical to manually tune parameters of algorithms such as PIV. There is growing interest in applying data-driven and machine-learning approaches to physics and materials discovery \citep{brunton2016,debezenac2019,cichos2020a,colen2021a,joshi2022b,zhou2021,brunton2022,golden2023,li2024}, but these approaches are limited by the availability of training data. The ability of DLOF to autonomously generate high-quality velocity fields is a crucial step for advancing these applications.

\section*{Acknowledgments}
This work was supported by the Department of Energy (DOE) DE-SC0022291. Computing resources were provided by the NSF XSEDE allocation TG-MCB090163 and the Brandeis HPCC which is partially supported by the NSF through DMR-MRSEC 2011846 and OAC-1920147. The authors thank Michael M. Norton (Physics Deparment, Brandeis University) for insightful feedback on the manuscript.

\section*{Supplementary Information}
In Experiment 1, active nematics were prepared as previously described~\citep{tayar2022}. The sample contained 1 mM ATP with 1 mg/mL Alexa-647 labeled MTs. We doped the active mixture with 10 ng/mL of Alexa-488 labeled MTs. We imaged these samples on an epi-fluorescence microscope (Nikon Ti2) equipped with an oil immersion objective 60X (NA 1.25) objective. The densely labeled Alexa-647 MTs and sparsely doped Alexa-488-labeled MTs were sequentially imaged every 1.50 seconds using a motorized fluorescence filter turret. 

In Experiment 2, active nematics were prepared following a similar procedure. The sample was mixed with 0.5 $\mu$m carboxyl beads which fluoresced at 488 nm. To suppress bead aggregation, the beads were coated with amine-PEG (20 KDa PEG) using a previously described protocol~\citep{garting2019}. The beads were mixed into the rest of the active mixture, which was then introduced into the sample chamber. The sample was spun down in a swinging bucket rotor (Sorvall LYNX 6000) at 2000$\times$g for 10 minutes. These samples were imaged with a Nikon Eclipse microscope and PCO Edge 4.2 camera using a 60X, 1.25 NA objective. The two channels (647 and 488 nm) were imaged sequentially, with an overall time step of 0.25 s, using a Lumencor Spectra light engine and a multi-band filter cube. 

For all active nematic datasets, PIV was performed using PIVLab v. 2.61 in MATLAB \citep{thielicke2021}. For pre-processing, CLAHE was applied with a window size of 32 pixels, and auto-contrast stretch was applied. For analyzing the PIV, 3 passes were used, with interrogation windows of 64, 32, and 16 pixels, and step sizes 32, 16, 8 pixels, respectively. For the sub-pixel estimator, the Gauss 2x3-point setting was used. The standard setting was used for correlation robustness. For post-processing the PIV fields, the vector validation routine with threshold of 8 times the standard deviation was used, with a local median threshold of 3. The routine was set to interpolate missing data. 

\bibliographystyle{apsrev4-1}
\bibliography{references,refs_from_joshi2023}
\bibliographystyle{rsc}
\end{document}